\documentclass[prd,aps,floats,twocolumn,nofootinbib]{revtex4-1}
\usepackage{slashed}
\usepackage{mathtools}
\usepackage{amsfonts}
\usepackage{amssymb}
\usepackage{epsfig}

\usepackage{bm}

\begin{document}

\newcommand{\m}[1]{\mathcal{#1}}
\newcommand{\nn}{\nonumber}
\newcommand{\ph}{\phantom}
\newcommand{\eps}{\epsilon}
\newcommand{\be}{\begin{equation}}
\newcommand{\ee}{\end{equation}}
\newcommand{\bea}{\begin{eqnarray}}
\newcommand{\eea}{\end{eqnarray}}
\newtheorem{conj}{Conjecture}

\newcommand{\plk}{\mathfrak{h}}


\title{Evolving laws and cosmological energy}
\date{}

\author{Jo\~{a}o Magueijo}
\email{j.magueijo@imperial.ac.uk}
\affiliation{Theoretical Physics Group, The Blackett Laboratory, Imperial College, Prince Consort Rd., London, SW7 2BZ, United Kingdom}

\begin{abstract}
We couple the issue of evolution in the laws of physics with that of violations of energy conservation. Avoiding a time dependence in terms of coordinate time, we define evolution as a function of time variables canonically dual to ``constants'' (such as $\Lambda$, the Planck mass or the gravitational coupling), mimicking a procedure associated with one formulation of unimodular gravity. We then introduce variability via a dependence of {\it other} fundamental ``constants'' on these clocks. 
Although this is not needed, sharper results are obtained if this procedure violates local Lorentz invariance, which we define in the spirit of Horava-Lifshitz theories (modifying a $3+1$ split action, so that a Lorentz invariant 4D reassembly is no longer possible).  
We find that variability in the ``laws of physics'' generically leads to violations of energy conservation if either a matter parameter varies as a function of a gravitational clock, or a gravity parameter depends on a matter clock, with the other combinations  sterile. In general the matter components associated with the varying parameter or the clock absorb or give off the violated energy. We illustrate this with a variety of clocks (ticking unimodular time, Ricci time, etc)
and parameters (mainly the gravitational and matter speed of light, but also $\Lambda$). 
We can accommodate in this construction (and improve on) several early Varying Speed of Light models, allowing for variability effects related to the spatial curvature and $\Lambda$ to cause creation of radiation and a Hot Big Bang. 
\end{abstract}

\maketitle

\section{Introduction}

John Wheeler reputedly stated that ``everything comes out of higgledy-piggledy'', to convey the view that the laws of Nature are subject to (possibly random) mutation, rather than being set in stone~\cite{PaulDavis}. In this paper we investigate whether such a mutability could be the origin of the matter and energy in our Universe. Symmetries and conservation laws are intimately connected (as enshrined in Noether's theorem); specifically the time-translation symmetry of the laws of physics implies conservation of matter and energy. This suggests a logical connection between Wheeler's view and cosmic creation out of nothing.
A major question hanging over this possibility is how to define the ``time in terms of which'' the laws evolve. The definition of time is closely related to the laws one accepts (to the extent that many propositions in any system are circular), so the construction of evolving laws and their associated times may indeed be higgledy-piggledy.

In this paper we take a rather conservative view on the matter (for more radical alternatives see, for example,~\cite{Smolin, Nielsen}). We propose interweaving two strands of thought. Along one strand, it has for long been known that one can frame evolution in the physical laws in terms of variability in the fundamental constants they employ. This goes back to at least Dirac's seminal paper~\cite{dirac} (see~\cite{anth} and references therein). Following another strand, one can obtain robust definitions of time by demoting the constants of nature to mere constants of motion (as done for the cosmological constant in unimodular gravity~\cite{unimod1,unimod,UnimodLee1,alan,daughton,sorkin1,sorkin2,Bombelli,UnimodLee2}, specifically according to the procedure of~\cite{unimod}). Employing the recipe in~\cite{unimod}, one finds that the canonical conjugates 
of the demoted physical constants are excellent candidates for physical, relational clocks, capable of surviving quantum gravity, among other blessings (see~\cite{gielen,gielen1}, for example). 

As an exploratory hypothesis, in this paper we propose a ``parting of constants'': perhaps the fate of some constants is to provide clocks via the procedure of~\cite{unimod1,unimod,UnimodLee1,alan,daughton,sorkin1,sorkin2,Bombelli,UnimodLee2}, whereas the fate of others is to vary in terms of such clocks. This orderly evolution might not have been to Wheeler's taste, but, as we will see, it will allow us a measure of pragmatic progress.

\section{A possible stage for evolution}\label{EvolutionSetup}
We consider variability within a set of target parameters $\bm\beta$, which could be any fundamental constant. The starting point is an action $S_0$ (which can be standard General Relativity or not) 
with all ``constants'' usually set to 1 restored, placed inside the space-time integral, and their different roles dissociated in the following sense.  For example, $c$, colloquially the ``speed of light'', plays several roles that are usually conflated, but which have no reason to be identified once we consider variability in ``$c$''.  This was stressed in~\cite{EllisVSL}, and applies to other ``constants'' demoted from their status: to give another example, Newton's constant $G$ has the double role of defining the Planck scale (appearing in the gravitational commutation relations) and of gravitational coupling to matter, and the two can be dissociated~\cite{vikman,JoaoLetter}. 
We may thus consider a target $\bm\beta$ which  includes $c_P$ and $G_P$ (appearing in the Planck scale, multiplying the gravity action), $c_g$ (the $c$ in the gravity metric), $c_m$ (the $c$ in the matter metric) and the coupling between matter and gravity, $G_M$.  We may then fix some of these, identify others, or impose a constraint between them. We could also consider further parameters, such as the electron charge $e$, and further dissociations of roles.

\subsection{The unimodular time prototype}\label{def-alpha}
As already stressed, 
the question then is: variability as a function of what? Rather than allowing a coordinate time $t$ to provide the brutal answer (thereby breaking time reparameterization invariance, as in~\cite{AM}), we propose the use of physical, ``relational'' time(s). These may be defined mimicking the Henneaux and Teitelboim (HT) formulation~\cite{unimod} of  unimodular gravity~\cite{unimod1,unimod,UnimodLee1,alan,daughton,sorkin1,sorkin2}, well known for producing a physical measure of time dual to the cosmological constant $\Lambda$: the so-called 4-volume or unimodular time~\cite{unimod,Bombelli,UnimodLee2}.

In the  HT formulation of unimodular theory
full diffeomorphism invariance is preserved (i.e. they are not restricted to volume preserving ones), but one adds to the base action $S_0$ an additional term:
\be\label{Utrick}
S_0\rightarrow S=S_0+S_U=
S_0- \int d^4 x \, \rho_\Lambda (\partial_\mu {\cal T}^\mu_\Lambda).
\ee
Here ${\cal T}^\mu_\Lambda$ is a  density, so that the added term is diffeomorphism invariant without the need of the metric or the connection.  Since the metric and connection do not appear in the new term, the Einstein equations (and other field equations) are left unchanged. 
In the standard theory $S_0$ does not depend on ${\cal T}^\mu_\Lambda$; hence, one equation of motion states the 
on-shell constancy of $\rho_\Lambda$. In addition, the zero-mode of the zero component of ${\cal T}^\mu_\Lambda$: 
\begin{equation}\label{TInt}
    T_\Lambda(\Sigma)\equiv \int_{\Sigma}  d^3 x\, {\cal T}^0_\Lambda
\end{equation}
(where $\Sigma$ is a spatial leaf within a 3+1 Hamiltonian splitting)
provides a definition of time, canonically dual to $\rho_\Lambda$.
For standard General Relativity we have:
\be
  T_\Lambda(\Sigma)=-\int^\Sigma_{\Sigma_0} d^4 x \sqrt{-g}\label{TLambda}
\ee
that is, minus the 4-volume between $\Sigma$ and a conventional ``zero-time'' $\Sigma_0$ leaf to its past, or unimodular time~\cite{unimod}. We will often omit the $\Sigma$ in our expressions. 

More generally~\cite{JoaoLetter,JoaoPaper}, we may select a set of $D$ constants $\bm\alpha$ and perform a similar exercise: 
\be\label{Utrickgen}
S_0\rightarrow S=S_0+S_U=S_0- \int d^4 x\, {\bm \alpha} \cdot \partial_\mu \mathbf{ \cal T}_{\bm \alpha}^\mu
\ee
where the dot denotes the Euclidean inner product in $D$ dimensional space. Just as it happens for unimodular theory, the zero-modes of the zero components of the density $\mathbf{ \cal T}_{\bm \alpha}^\mu$ provide definitions of time $\bm T_{\bm\alpha}$, dual to the on-shell constants $\bm\alpha$. Besides including $\rho_\Lambda$ in $\bm\alpha$ (as in unimodular theory) we can consider a more general setting, including the ``sequester''~\cite{pad,pad1}, where $c_P^2/G_P$ (responsible for the Planck mass) is an element of $\bm \alpha$, its canonical conjugate providing a Ricci time clock. 

\subsection{Evolution potentials}
We define ``evolution in the laws of physics'' by making the parameters ${\bm \beta}$ vary according to specified functions of the relational times $\bm T_{\bm\alpha}$: 
\begin{equation}
    \bm\beta=\bm\beta(\bm T_{\bm\alpha}).
\end{equation} 
If $\bm\beta$ overlap with $\bm\alpha$ this amounts to imposing a second-class constraint, but we do not consider this situation here. (The implications of such constraints vary from a technical nuissance to downright inconsistency, depending on the case; this is the reason for our ``parting'' the constants into 2 classes.) Instead, we investigate a sub-class of theories where {\it some} constants (the $\bm \alpha$)
are deconstantized so that their duals provide clocks, $\bm T_{\bm\alpha}$; and where a {\it distinct set} of ``constants'' (the $\bm \beta$) are allowed to  
vary as functions of the physical clocks provided by the former.

There are many combinations and choices for $\bm\alpha$ and $\bm\beta$ (to be explored in this paper and in a sequel), but given our emphasis on Lorentz invariance violation (LLIV) in this paper we will take as a starting point the selection (to be changed later): 
\begin{eqnarray}
     {\bm\beta}&=&\left(c_g^2,c_m^2,...
    \right)\\
    {\bm \alpha}&=&\left(\rho_\Lambda, \frac{3 c_P^2}{8\pi G_P},\frac{G_M}{G_P},... \right),\label{alpha}
\end{eqnarray}
where the extra factor of 6 in $\alpha_2$ is included for later convenience. 
For standard General Relativity:
\be\label{S0def}
S_0=\int d^4 x\, \sqrt{-g}\left[\frac{\alpha_2}{6}R+\alpha_3{\cal L}_M -\rho_\Lambda\right],
\ee
and the expressions for $T_{\bm\alpha}$ are well-known. Beside $T_1\equiv T_\Lambda$, we have
\begin{eqnarray}
    T_2\equiv T_R&=&\frac{1}{6}\int d^4 x \sqrt{-g} R\label{TRicci}\\
     T_3\equiv T_N&=&\int d^4 x \sqrt{-g} {\cal L}\label{TMatter}
_M\end{eqnarray}
i.e.: Ricci and Newton\footnote{Since $T_3$ is dual to Newton's $G_N$ we propose this nomenclature.} times
(see~\cite{pad,pad1,lomb,vikman,vikman1} in addition to~\cite{unimod1,unimod,UnimodLee1,alan,daughton,sorkin1,sorkin2,Bombelli,UnimodLee2}). 
$T_\Lambda$ and $T_R$ are nothing but the ``fluxes'' used in the sequester model~\cite{JoaoPaper,JoaoLetter,pad,pad1}.
We will also toy with $c_m$ being included in the $\bm \alpha$ rather than $\bm\beta$ and other combinations. 

The evolution functions 
$\bm\beta(\bm T_{\bm\alpha})$ are ``givens'', in the same sense that classical potentials in field theory are givens. This is supported by analogy, as we shall see in Sec.~\ref{masslessphi}. If LLIV is present, the $\bm \beta$ functions can be seen as ``LLIV-potentials''.  
In most cases we will assume they are step functions, modelling abrupt phase transitions, just as in~\cite{MoffatVSL,AM,VSLreview}.

\subsection{Gauge-invariance}
The unimodular-like extensions do not add new local degrees of freedom. This was remarked in~\cite{unimod} for $\Lambda$ (see also~\cite{vikaxion,unimodProca}) and is a general feature of all extensions (\ref{Utrickgen}). It  follows from the invariance of $S_U$ under 
the gauge symmetry:
\begin{equation}
    {\cal T}_\Lambda^\mu\rightarrow {\cal T}_\Lambda^\mu+\epsilon^\mu\quad {\rm with}\quad \partial_\mu\epsilon^\mu =0.
\end{equation}
It leads to first class constraints cancelling out the apparent new local degrees of freedom. 
However, the zero-mode of the zeroth component of ${\cal T}^0_\Lambda$ given by (\ref{TInt})
is gauge-invariant, so globally (and in the minisuperspace approximation) we do have an extra degree of freedom. 
For example, in standard General Relativity with a pure $\Lambda$ we have no degrees of freedom under homogeneity and isotropy (de Sitter with the pre-given $\Lambda$ is the only solution), whereas in unimodular theory we can specify the value of $\Lambda$.

Our $\bm \beta$, therefore, do not break gauge-invariance.  
Had we made the $\bm \beta$ (now functions of space as well) depend on the local ${\cal T}^0({\mathbf{x}})$ (and not just its zero mode) we would have broken the gauge symmetry. This is interesting (and was investigated in a different context in~\cite{unimodProca}), but it will not be considered here.
Finally, we note that,
although the gauge-invariant zero mode depends on the foliation $\Sigma$, the theory does not need to break local Lorentz invariance. However, it can do so, and the presence of a preferred $\Sigma$ may be a motivation for doing so. We now explain what we mean by this, and how it relates to the choice of $\bm\beta$. 


\section{Breaking local Lorentz invariance}\label{LLIV}
We define local Lorentz invariance violation (LLIV) in the spirit of Horava-Lifshitz (HL) theory~\cite{HL}. The idea is to take a local Lorentz invariant  action,  choose a foliation $\Sigma$, perform a 3+1 split as in the Hamiltonian formulation, and then change an aspect of the split action so that a local Lorentz invariant 4D reassembly is no longer possible. The kinetic term of HL theory is an example of this. One considers a generic kinetic term of the form~\cite{HL}:
\begin{equation}
    S_K=\frac{c^3}{16\pi G}\int dt\, d^3x \, \sqrt{|g|}[K_{ij}K^{ij}-\lambda (K_i^i)^2],
\end{equation}
so that when $\lambda=1$ it reduces to the kinetic term of General Relativity, and so complies with local Lorentz invariance. One then allows
$\lambda\neq 1$, so that the theory no longer derives from the  Einstein-Hilbert action. Although the action is still invariant under foliation-preserving (3D) diffeomorphisms and time reparameterization, it lost full 4D diffeomorphism invariance. This is merely an illustration of the process we have in mind, and we ignore other aspects of HL theory, such as detailed balance.

\subsection{A gravitational example}
In the same spirit, we 
take the Einstein-Cartan action subject to the operations described at the start Section~\ref{EvolutionSetup}:
\begin{equation}\label{ECaction}   S_{EC}=\int\frac{c_P^2}{32\pi G_P}  \epsilon_{ABCD}e^Ae^BR^{CD} [\Gamma]\end{equation}
(here $e^A$ is the tetrad and $\Gamma^A_{\;\; B}$ is the spin-connection). We then split it as~\cite{Thiemann}:
\begin{eqnarray}\label{ECaction3+1}
S_{EC}&=&
\int dt\, d^3x\, \frac{c_P^2}{16\pi G_P}[2\dot{K}^i_aE^a_i-
(NH+N^aH_a+ \nn\\
&&+ N^{ij}G_{ij})],
\end{eqnarray}
where $K^i_a$ is the extrinsic curvature connection\footnote{From which the Ashtekar connection can be built by a canonical transformation~\cite{Thiemann}, should we wish to generalize the theory we are about to propose.}, $E^a_i$ is the densitized inverse triad, and the last three terms are the Hamiltonian, Diffeomorphism and Gauss constraints:
\begin{eqnarray}
H&=&\frac{1}{\sqrt{h}}(K^j_a K^i_b-K^i_a K^j_b )E^a_i E^b_j-c_g^2\sqrt{h} R_3\label{Ham-const}\\
H_a&=&-2D_b(K^i_a E^b_i -\delta^a_b K^i_c E^c_i)\\
G_{ij}&=&K_{a[j} E^a_{k]}
\end{eqnarray}
(where $R_3$ is the 3D curvature, $h$ is the determinant of the 3 metric and $D_a$ the 3D covariant derivative)
enforced by corresponding Lagrange multipliers $N$, $N^a$ and $N^{ij}$. In writing (\ref{ECaction3+1}) we have chosen units as follows:
the coordinates satisfy $[x^0]=T$ and $[x^i]=L$; 
the metric is dimensionless 
and so $[e^A_\mu] =[E^a]=[N]=[N^i]=L^0$; the extrinsic curvature has units of inverse time squared $[K^a_b]=1/T^2$ (derivatives with respect to time), but the 3D curvature has units of inverse length squared, $[R_3]=1/L^2$; the volume element is written in terms of time and 3D spatial volume. With these choices the pre-factor of the action is proportional to $c_P^2/G_P$, as written in (\ref{ECaction3+1}), and  a factor of $c_g^2$ appears in the last term of the Hamiltonian constraint (\ref{Ham-const}) (with the Hamiltonian constraint having units of $[H]=1/T^2$). 

If $c_g$ is constant none of this matters. But if we allow its variation (even if only as a function of a global time variable, $c_g=c_g(\bm T_{\bm\alpha})$, as in Section~\ref{EvolutionSetup}), then we have an analogy with the kinetic term in HL for $\lambda\neq 1$. Although we started from  (\ref{ECaction}), once we allow $c_g=c_g(\bm T_{\bm\alpha})$ the split action  (\ref{ECaction3+1}) can no longer arise from (\ref{ECaction}). Any attempt to do so would produce extra terms in the derivatives of $c_g$.
Since the connection $\Gamma^A_{\;\; B}$ has units of $1/T$, the extra terms in the action arise from:
\begin{equation}
    \int \epsilon_{ABCD}e^Ae^B \Delta R^{CD} 
\end{equation}
where:
\begin{equation}
    \Delta R^A_{Bi}=\partial_\mu \frac{1}{c_g}\Gamma^A_{Bi}dx^\mu dx^i.
\end{equation}
These terms break local Lorentz invariance.
Hence the action (\ref{ECaction3+1})
obtained by neglecting these terms when starting from the LI Eq.~(\ref{ECaction}) must be LLIV.



The above choice of units is crucial to LLIV. If $K^i_a$ and $R_3$ are defined with the same units, then an overall common factor of $c_g$ appears in (\ref{Ham-const}), so that the time variations in $c_g$ can be absorbed into a redefinition of $N$. Also, if we define time as $x^0$, that is time with units of length, no extra terms appear going from (\ref{ECaction}) to (\ref{ECaction3+1}) (but see~\cite{CovVSL}). With such choices, in effect the notation has absorbed the assumption of a fixed speed of light, with time and space rendered equivalent and the speed of light losing its meaning. 

Upon quantization the gravitational commutation relations inferred from 
(\ref{ECaction3+1}) are:
\begin{eqnarray}\label{commAsh}
[K^i_a(\boldsymbol{x}),E_j^b(\boldsymbol{y})]&=&i\frac{8\pi G_P\hbar}{3 c_P^2}
\delta^b_a\delta^i_j\delta(\boldsymbol{x}-\boldsymbol{y})\nn\\
&=&
il_P^2 c_P \delta^b_a\delta^i_j\delta(\boldsymbol{x}-\boldsymbol{y}),
\end{eqnarray}
where $l_P^2=8\pi G_P\hbar/ c_P^3$ is the Planck length
(note that the delta function has units of $1/L^3$, so it all matches). This justifies the statement that $G_P$ and $c_P$ are the relevant variables setting the Planck scale. Indeed this ``speed of light'', $c_P$, is nothing but a conversion factor between different parameterizations of the Planck scale with different dimensions, and it will be absorbed into a single parameter ($\alpha_2$, as in Eq.(\ref{alpha})), and ignored for the rest of this paper. 

\subsection{A matter example}\label{LLIVmatter}
Theories where $\alpha_3=G_N/G_P$ and its dual, $T_N$, come into play offer another example of LLIV in the same spirit (see (\ref{alpha}) and (\ref{S0def}) for definitions). Integrations by parts in ${\cal L}_M$, which are innocuous in the standard theory, lead to different theories if $\alpha_3$ is included and there is $T_N$ dependence. This induces LLIV if the integration by parts is designed to do so.  

Let us consider extension (\ref{Utrickgen}) applied to $\alpha_3$, and a foliation $\Sigma_t$ used to define the associated gauge-invariant $T_N$ (as in (\ref{TInt})). The 3+1 split of this term in $S_U$ is:
\begin{equation}\label{a3dotTn}
    S=S_0-\int dt\, \alpha_3(t)\dot T_N
\end{equation}
where we have {\it not} assumed homogeneity. If there is $T_N$ dependence in $S_0$, then $\alpha_3$ will change in time {\it but not in space} in this foliation, according to the corresponding Hamilton equation, $\dot\alpha_3=\partial H/\partial T_N$. 

Now take a massless scalar field in flat spacetime as an example. Upon a 3+1 split in the same foliation its action becomes:
\begin{equation}\label{Sphi1}
    S_\phi=\int d^3x\,dt \left(\dot\phi \Pi_\phi  -\frac{N}{2}\left(\Pi_\phi ^2 + (\partial_i\phi)^2
\right)\right)
\end{equation}
and we could perform an integration by parts in time in the first term to bring this to:
\begin{equation}\label{Sphi2}
    S_\phi=\int d^3x\,dt \left(\dot \Pi_\phi \phi  -\frac{N}{2}\left(\Pi_\phi ^2 + (\partial_i\phi)^2
\right)\right)
\end{equation}
(with a $\Pi_\phi\rightarrow-\Pi_\phi$ redefinition). Both  actions are (at least classically) equivalent if $\alpha_3$, as defined in (\ref{S0def}), is not deconstantized. Both lead to a massless Klein-Gordon equation. 
Under (\ref{a3dotTn}), however, it matters if in (\ref{S0def}) we take (\ref{Sphi1}) or (\ref{Sphi2}) for ${\cal L}_M$. If $S_0$ depends on $T_N$, so that $\alpha_3=\alpha_3(t)$ in this foliation, they are different theories. The first is still Lorentz invariant and has equation of motion $\partial_\mu (\alpha_3 \partial^\mu \phi)=0$. The second has LLIV, and satisfies:
\begin{equation}
    -\alpha_3(t) \ddot \phi +\nabla \phi=0
\end{equation}
in this foliation. Thus, $\alpha_3$ theories may be used to implement LLIV in the matter sector, even though they do not have to. 

Note that even when $\alpha_3$ is a constant, the on-shell value of the action is different in the two theories. In the first, the on-shell Lagrangian is the pressure (should there be a potential, the Lagrangian would be $p=K-V$). In the second, it is minus the energy density (minus the Hamiltonian, since the canonical pair terms vanish).  This affects the on-shell expression for $T_N$.

\section{Reduction to homogeneity and isotropy}\label{MSSVSL}
Reducing the action $S=S_0+S_U=S_G+S_M +S_U$ (where $S_G$ and $S_M$ are the gravity and the matter base actions) to homogeneity and isotropy we find:
\begin{eqnarray}
S_{G}&=& V_c\int 
dt\, 
\alpha_2
\bigg(\dot{b}a^2 + Na (b^{2}+k c^2_g)\bigg)\label{SGMSS}\\
S_M&=&V_c\int dt \,
\alpha_3
\bigg(\dot m_i \phi_i -Na^3(\rho_\Lambda +\rho_M)\bigg)\\
S_U&=&V_c\int dt\, \dot {\bm \alpha}\cdot {\bm T} =  V_c\int dt\, \dot \rho_\Lambda T_\Lambda +... \, .\label{SUMSS}
\end{eqnarray}
Here $a$ is the expansion factor, $b$ is the minisuperspace connection variable (an off-shell version of the Hubble parameter, since $b= \dot a$ on-shell, if there is no torsion), $k$ is the spatial curvature, $N$ is the lapse function and $V_c=\int d^3 x$ is the comoving volume of the region under study, assumed finite throughout this paper (in the quantum cosmology classical literature one usually chooses $k=1$ and $V_c=2\pi^2$; see~\cite{Afshordi} for a discussion of the criteria for the choice of $V_c$). We allow for a matter action $S_M$ with multi-fluids and possibly scalar fields (see later applications), with $\rho_M=\sum_i\rho_i$ with 
\begin{equation}
    \rho_i=\frac{m_{i}}{a^{3(1+w_i)}}
\end{equation}
with $w_i$ the fluid's equation of state (here assumed constant), $m_i$ a constant of motion in the usual theory, dual to variables $\phi_i$ as defined in the Lagrangian formulation of perfect fluids of~\cite{brown}, reduced to MSS as in~\cite{gielen,gielen1,twot}. We will  generalize this in Section~\ref{cmingamma}.


For this action, we therefore have total Hamiltonian:
\begin{equation}
    H=NaV_c \bigg[-\alpha_2 (b^{2}+k c_g^2
)+ \alpha_3 ( \rho_\Lambda +\rho_M)a^2\bigg]
\end{equation}
and Poisson brackets, 
\begin{eqnarray}
    \{b,A^2\}&=&\frac{1}{V_c}\\
     \{\alpha_i,T_i\}&=&\frac{1}{V_c}\label{PBMSS2}
\end{eqnarray}
with the conjugate of $b$ given by
\be\label{A2a2}
A^2=a^2 \alpha_2. 
\ee
Upon quantization the gravitational commutation relation in MSS can be written as:
\begin{eqnarray}\label{commutator}
   \left[\hat b,\hat{ a^2}\right]&=&i\frac{8\pi G_P\hbar}{3V_c c_P^2}=\frac{i l_P^2c_P}{3 V_c} \equiv i\plk c_p
\end{eqnarray}
which is a reduced version of (\ref{commAsh}). The Planck constant $\plk$ is the relevant quantity for minisuperspace quantum cosmology.

We first assume that $\alpha_2$ is kept constant (this leads to a fork in the formalism, as we will investigate in Section~\ref{alpha2}). Then $a$ and $A$ can be used interchangeably. In finding the Poisson equations, we stress that, as per the usual definition of the Poisson bracket, $\bm \alpha$ and $\bm T$ (and so also $\bm \beta$) are kept fixed when evaluating Poisson brackets involving $b$ and $a$. As a result their equations of motion are obtained from the usual ones simply by inserting varying constants where usually they appear as fixed parameters. Hence, for all theories based on this set up (with the exclusion of those with varying $\alpha_2$, as we will see),  
we have a Hamilton constraint (obtained by varying with respect to $N$) and Poisson equations for for $a$ and $b$:
\begin{eqnarray}
    b^{2}+k c^2_g&=&\frac{\alpha_3}{\alpha_2} \rho a^2\nn\\
    \dot a&=&\{a,H\}=Nb\nn\\
    \dot b&=&\{b,H\}=-\frac{\alpha_3}{2\alpha_2} (\rho+3p )Na
    .\label{EOM}
\end{eqnarray}
with 
\bea
\rho&=&\rho_\Lambda + \rho_M=\rho_\Lambda + \sum _i \rho_i\\
p&=&p_\Lambda + p_M=- \rho _\Lambda + \sum _i w_i
\rho_i.\eea
This is equivalent to directly inserting varying constants into the unmodified Friedman and Raychaudhuri equations, as in~\cite{AM}:
\begin{eqnarray}
  \dot a^2 + k c^2_g&=&\frac{\alpha_3}{\alpha_2} \rho a^2\nn\\
  \ddot a&=&-\frac{\alpha_3}{\alpha_2} ( \rho +3  p)]a.\nn
\end{eqnarray}
(in the $N=1$ gauge for simplicity). Assuming further a fixed $\alpha_3$ (not necessarily equal to 1), by 
dotting the first equation and comparing with the second,  a consistency relation can be obtained:
\begin{equation}\label{non-cons}
    \dot { \rho}+3\frac{\dot a}{a}(\rho+ p)=
    \frac{k\alpha_2}{a^2\alpha_3}\frac{dc_g^2}{dt}
\end{equation}
(valid in any gauge). This the central Eq.(5) in~\cite{AM}:
\begin{equation}\label{non-consAM}
    \dot { \rho}+3\frac{\dot a}{a}(\rho+ p)=
   \frac{3kc_g^2}{4\pi G_M a^2}\frac{\dot c_g}{c_g}
\end{equation}
given that we fixed $c_P$ (together with $\alpha_2$) so the barred densities in~\cite{AM} there can be dropped. 

It is interesting that the most abstruse assumptions in~\cite{AM} are in fact natural from the point of view of the construction in Section~\ref{EvolutionSetup}. In~\cite{AM} it was postulated that varying constants did not produce stress-energy-momentum $T_{\mu\nu}$, or changed the Einstein equations, something undoubtedly odd, say, from the perspective of scalar-tensor theories. This is just what happens here for two reasons. First, the unimodular-like procedure designed in Section~\ref{def-alpha} for producing clocks is such that the metric does not enter $S_U$, so no $T_{\mu\nu}$ 
is produced. Second the theory has a Hamiltonian structure, so that the Poisson equations for one set of variables are obtained fixing all other variables. Hence the Einstein equations in their 3+1 split form are obtained fixing all other phase space variables including the time variables ${\bm T}_{\bm\alpha}$
and by extension the varying constants $\bm\beta$. 

We add that this is not a miracle and in fact it is not true in general. If we do not fix $\alpha_2$, then new terms do appear in the Einstein equations, unless we go beyond minimal assumptions regarding the sympletic/Hamiltonian structure of the theory, as we will see in Section~\ref{alpha2}. In order to avoid extra terms under a varying $\alpha_2$ one would need to fix the $\alpha_2$ multiplying first term in (\ref{SGMSS}), $\dot b a^2$, whilst allowing deconstantization for the $\alpha_2$ multiplying the second term and the gravitational Hamiltonian. Such a procedure goes beyond Section~\ref{alpha2} and clearly requires LLIV.






\section{Matter-based clocks}\label{firstmodels}
We thus justify the central equations of~\cite{AM}, 
but we can go beyond. A major weakness of~\cite{AM} was an absence of information on how the energy conservation violating 
last term in (\ref{non-cons}) would be shared between the various matter components (including $\Lambda$). This was not an issue in scenarios where only radiation is present, as envisaged in~\cite{AM}. It turns out we can be more predictive in this respect, opening up the doors to other scenarios. As a first rule of thumb, this energy is dumped into (or extracted from) whatever matter form is conjugate to the relational time variables upon which $c_g$ depends (should the clocks be ``matter clocks'', as we now illustrate. Throughout this Section we fix $\alpha_2$ and $\alpha_3$ (see next Section for the associated issues). 

\subsection{Pure unimodular varying speed of light (VSL)}\label{unimodVSL}
The simplest theory follows from $\bm \alpha=\rho_\Lambda$ (pure unimodular $S_U$) and $\bm\beta =c_g$, so that $c_g=c_g(T_\Lambda)$. 
Then, beside Equations (\ref{EOM}) (which always apply), we have two more Hamilton equations:
\begin{eqnarray}
   \dot T_\Lambda &=&\{T_\Lambda,H\}=-N\alpha_3 a^3\label{unimodvsl1}\\
    \dot \rho_\Lambda&=&\{\rho_\Lambda,H\}=\frac{\partial H}{\partial T_\Lambda}=   -Na\alpha_2k\frac{d c_g^2}{dT_\Lambda}
  .\label{unimodvsl2}
\end{eqnarray}
The first equation just states that $T_\Lambda$ is the 4-volume time if $\alpha_3=1$ and agrees with (\ref{TLambda}) modulo a factor of $V_c$. By introducing $V_c$ in Poisson bracket (\ref{PBMSS2}) we are defining {\it  intensive} times, i.e. time per unit of spatial volume. In this case we are defining time as the 4-volume per unit of 3-volume, so that:
\be
T_\Lambda=-\alpha_3\int dt N \, a^3,
\ee
with units of time. 
The time dependence of $c_g$ does not affect this equation, since it only cares about the  Hamiltonian dependence on $\rho_\Lambda$.

The second equation implies that the energy source/sink term contained in (\ref{non-cons}) goes fully into the vacuum energy, since it implies (when combined with (\ref{unimodvsl1})):
\begin{equation}\label{nonconsLambda}
      \dot { \rho}_\Lambda+3\frac{\dot a}{a}( \rho_\Lambda+ p_\Lambda)=\dot{\rho}_\Lambda= 
        \frac{k\alpha_2}{a^2\alpha_3}\frac{dc_g^2}{dt}.
\end{equation}
This is true even if there are other forms of matter $\rho_i$ and is the simplest example of a pattern which we now uncover by generalization: the violations of energy conservation are absorbed by the matter ingredient dual to the clock. 



\subsection{The matter speed of light, $c_m$}\label{cmingamma}
Turning now to the matter speed of light, $c_m$, we note that it can vary (or be fixed) independently from $c_g$ (e.g.~\cite{EllisVSL,Coleman}). For definiteness, we model matter as a mixture of isentropic fluids according to~\cite{brown,gielen,gielen1,twot}. Imposing  homogeneity and isotropy the action becomes:
\begin{eqnarray}\label{brown}
     S_M&=&V_c\int dt\sum_i \left(a^3n_i\dot \Theta_i -N a^3 \rho_i(n_i)\right) .
\end{eqnarray}
For each specie $i$, $\Theta_i$ is a Lagrange multiplier, so that $n_i$ is the (spatial) volume density of a conserved particle number (i.e.: $\Pi_i=n_ia^3$ is conserved). Thermodynamics shows that the chemical potential and pressure are given by:
\begin{eqnarray}
    \mu_i&=&\frac{p_i+\rho_i}{n_i}=
    \frac{\partial \rho_i}{\partial n_i}
    \label{mudef}\\
    p_i&=&n_i\frac{\partial\rho_i}{\partial n_i}-\rho_i.\label{pdef}
\end{eqnarray}
Restoring $c_m$ we have in general:
\begin{equation}
    \rho_i=\rho_i(n_i,c_m), 
\end{equation}
so that:
\begin{equation}\label{mucm}
    \mu_{ic}=\frac{\partial\rho_i}{\partial c_m^2}
\end{equation}
can be thought of as the speed of light ``chemical'' potentials, in analogy with (\ref{mudef}). 
For example, for dust $\rho=nm_0(c_m)c_m^2$ (with $m_0$ a constant in some cases); for black body radiation $\rho=n^{4/3}\hbar c_m$. The $c_m$ chemical potentials are a property of each matter specie just like $w_i$. Thus, detailed particle physics enters the cosmological argument\footnote{Obviously we could complicate the model by attributing different $c_{mi}$ to different species as in~\cite{Carroll}.}.  


If we include $c_m^2$ in $\bm \alpha$, then (\ref{brown}) becomes:
    \begin{equation}\label{SMcm2}
      S_M=V_c\int dt \left[\dot c_m^2 T_c+\sum_i\left( \Pi_i\dot  \Theta_{ i} -N a^3 \rho_i\left(\frac{\Pi_i}{a^3},c_m^2\right)\right) \right]. 
\end{equation}
If $w_i\neq -1$ is constant, then  $\rho_i= f_i n_i^{1+w_i}$, where $f_i$ is a dimensionful proportionality constant~\cite{brown}. Hence:
\begin{equation}
    \rho_i=n_i^{1+w_i}f_i(c_m^2)
\end{equation}
and we can perform a canonical transformation from $\{\Pi_i,\Theta_i\}$ to $\{m_i,T_{mi}\}$ so that:
\begin{equation}\label{SMcm1}
      S_M=V_c\int dt \left[\dot c_m^2 T_c+\sum_i\left(\dot m_i   T_{mi} -N \frac{m_if_i(c_m^2)}{a^{3w_i}}
      \right) \right]
\end{equation}
leading to the time formula:
\begin{eqnarray}
    \dot T_c&=&-N\sum_i\frac{m_if_i'}{a^{3w_i}}.
    \label{Tcm}
\end{eqnarray}
For a model 
where $c_g\in \beta$ only depends on $T_c$ we have:
\bea
\dot c_m^2&=&\frac{\partial H}{\partial T_c}=
-Na \alpha_2k\frac{\partial c_g^2}{\partial  T_c}
\eea
resulting in:
\begin{eqnarray}
     \dot \rho_i+3\frac{\dot a}{a}(\rho_i+p_i)
     &=&
     \frac{m_i\dot f_i}{a^{3(1+w_i)}}=
\frac{m_i f'_i}{a^{3(1+w_i)}}
     \frac{dc_m^2}{dt}
     \nn\\
&=&
\frac{k\alpha_2}{a^2}\frac{-\frac{Nf_i' m_i}{a^{3w_i}}}{\dot T_c} \frac{dc_g^2}{dt}
\label{dotrhocgtc}
\end{eqnarray}
which summed over $i$ leads to 
Eq.~(\ref{non-cons}), as it should. But we have learnt more: in this theory the more a specie depends on $c_m$ the more it is affected by energy conservation violations, an extension of the pattern in the previous subsection.

We could instead have $c_m\in \bm\beta$, 
and in particularly set $c=c_g=c_m$. We could also write $\rho_\Lambda=m_\Lambda f_\Lambda(c_m^2)$, and take $m_\Lambda$ for $\bm\alpha$ leading to a slightly modified unimodular clock, with $\dot T_\Lambda=-Na^3 f_\Lambda'$. Then:
\begin{eqnarray}\label{dotLcm}
     \dot \rho_\Lambda=&
     (m_\Lambda f_\Lambda)^.=
\dot m_\Lambda f_\Lambda 
+ m_\Lambda f'_\Lambda
     \frac{dc_m^2}{dt},
\end{eqnarray}
whereas for the other matter components ($i\neq \Lambda$):
\begin{eqnarray}
     \dot \rho_i+3\frac{\dot a}{a}(\rho_i+p_i)
     &=&
     \frac{m_i\dot f_i}{a^{3(1+w_i)}}=
\frac{m_i f'_i}{a^{3(1+w_i)}}
     \frac{dc_m^2}{dt}.
     \nn  
\end{eqnarray}
The extra terms in $\dot c_m$ are $\partial{\bm\beta}/\partial {\bm T}_{\bm\alpha}$ terms. 
The first term in (\ref{dotLcm}) can be expanded using:
\begin{eqnarray}
      \dot m_\Lambda &=&\frac{\partial H}{\partial T_\Lambda }=\frac{\partial H_g}{\partial T_\Lambda }+\frac{\partial H_m}{\partial T_\Lambda }\nn\\
      &=&-Nak\alpha_2 \frac{\partial c_g^2}{\partial T_\Lambda }+Na^3\sum_i \frac{\partial\rho_i}{\partial c_m^2}\frac{\partial c_m^2}{\partial T_\Lambda }
\end{eqnarray}
so we find after some algebra:
\begin{equation}
      \dot \rho_\Lambda
     =\left(\frac{k\alpha_2}{a^2}
     -\sum_{i\neq \Lambda} 
     \frac{m_i f'_i}{a^{3(1+w_i)}}
     \right)\frac{dc^2}{dt},
\end{equation}
(where we have set $c_m=c_g$).
Again, summing over all the $i$ we get (\ref{non-cons}). 
Even though the net production of matter is not affected by $c_m(T_\Lambda)$, the energy equations for each specie receive a source/sink term dependent on their $c_m$ chemical potentials. If $c_m$ is part of $\bm\beta$ it can be seen as an interaction potential between the different components, allowing for interchange of matter.

\subsection{An analogy}
\label{masslessphi}
We could use the fluids as clocks ($m_i\in\bm\alpha$ with conjugate $T_{mi}$ ticking ``temperature time'' for radiation, proper time for dust particles, for example) and likewise with any scalar field (as was done in~\cite{gielen,gielen1}). We will not explore this option in this paper because it is less fundamental than targeting proper constants; however here we draw attention to an interesting analogy. Take a massless scalar field:
\begin{equation}
    S_\phi=V_c\int dt \left(\dot\phi \Pi_\phi  -N \frac{\Pi_\phi ^2}{2 a^{3 }}\right),
\end{equation}
so the field $\phi$ and its conjugate $\Pi_\phi$ satisfy:
\begin{eqnarray}
    \dot T_\phi\equiv \dot \phi &=&N\frac{\Pi_\phi }{a^{3 }}\\
    \dot \Pi_\phi &=&-\frac{\partial H}{\partial \phi }.
\end{eqnarray}
It then looks as if we can use $\Pi_\phi$ as a constant $\bm\alpha$ and $\phi$ as its conjugate time, but this is at best an effective description. Nonetheless if for example we set $c_g=c_g(T_\phi)$ we find that the non-conservation equation for $\phi$ can be written as:   
\begin{equation}
    \ddot \phi +3\frac{\dot a}{a} \dot\phi=
    \alpha_2 k \frac{\partial c_g^2}{\partial T_\phi}.
\end{equation}
since $\rho_\phi=p_\phi=\dot \phi^2/2$. It suggests interpreting: 
\begin{equation}
    U_{LLIV}=- \frac{3kc_P^2 c_g^2(\phi)}{8\pi G_P } 
\end{equation}
as a LLIV potential, source of energy conservation violations. It makes sense to look at $\bm\beta(\bm T_{\bm\alpha})$ as ``potentials''.

\section{Gravity clocks}\label{formaldevs}
The examples in the Section~\ref{firstmodels}  show a preliminary pattern. Suppose a physical clock is made of matter: it has a matter constant of motion, which is the canonical dual variable to its variable ``moving hands''. If the laws of nature depend on the time ticked by this clock via a gravitational parameter ($c_g$ in our example), then there are energy conservation violations.
This surplus/deficit of energy goes into the clock's constant of motion, and so into the clock system. We saw this happen if the clock is unimodular time, with $\Lambda$ collecting or giving off the energy. 
In the case of a $c_m$ clock, the dual constant of motion is shared by several fluids, with the total energy violation apportioned between the matter components depending on how much each contributes to the clock. 

However, we already saw this pattern break down if a matter parameter, instead, depends on a matter clock (as with $c_m(T_\Lambda$)).
The pattern also fails to apply to gravitational parameters depending on gravitational clocks. To explore this, we need to develop formalism for $T_R$ and $T_N$, bringing $\alpha_2$ and $\alpha_3$ into the calculation.

\subsection{LLIV theories with 
Ricci and Newton clocks}
\label{alpha2}
The calculation in Section~\ref{MSSVSL} is different if the Planck scale $\alpha_2$ is one of the $\bm \alpha$ and the $\bm \beta$ depend on its associated time (Ricci time, $T_R$, see (\ref{TRicci})).
Then, it is $A$ (as defined in (\ref{A2a2})) and not $a$ that should be used when computing Poisson brackets. It is useful to write the Hamiltonian as a function of $A$ instead of $a$ (or simply to evaluate the Euler-Lagrange equations, in this case). The
equations of motion are:
\begin{eqnarray}
    b^{2}+k c^2_g&=&\frac{\alpha_3}{\alpha_2} \rho a^2\label{hamG}\\
    \dot a+\frac{\dot \alpha_2}{2\alpha_2}a&=& Nb\label{dotaG}\\
    \dot b&=& -\frac{\alpha_3}{2\alpha_2 } (\rho+3p) Na\label{dotbG}
\end{eqnarray}
and the central assumption in~\cite{AM} is violated (a new term appears in the second equation). We remark that for $\alpha_3=1$ these equations of motion are the same as those for a non-dynamical scalar field in the Jordan frame (i.e. Brans-Dicke theory in the first order formulation with $\omega_{BD}=0$) as studied in~\cite{flanagan,vollick,vollick2}\footnote{We can also rephrase the new term in the equations as a theory with torsion, due to a non-fixed Lambda potential, (as in~\cite{Lee2,tom}, in this case with the even-parity torsion given by $T=\dot\alpha/(2N\alpha)$). The crucial difference here is the unimodular addition, and the fact that the potentials are non-local functions of the conjugate of the dilaton, rather than of the dilaton itself.}.

A non-conservation equation can be obtained as before, dotting the Hamilton constraint (\ref{hamG}), and using  (\ref{dotaG}) and (\ref{dotbG}) to eliminate $b$ and $\dot b$. It may then be convenient to define densities converted into geometrical quantitities (with units of $1/T^2$):
\begin{equation}
    \tilde \rho_i=\frac{\alpha_3}{\alpha_2}\rho_i
\end{equation}
(and likewise for $p$), to find:
\begin{equation}
    \dot {\tilde \rho}+3\frac{\dot a}{a}({\tilde \rho}+\Tilde{p})=-\frac{\dot\alpha_2}{\alpha_2}\frac{{\tilde \rho}+3\tilde p}{2}+\frac{k}{a^2}\frac{dc_g^2}{dt}.
\end{equation}
Reverting to the original variables, we have:
\begin{equation}
    \dot \rho+3\frac{\dot a}{a}(\rho+p)=-\frac{\dot \alpha_3}{\alpha_3}\rho +\frac{\dot\alpha_2}{\alpha_2}\frac{\rho-3p}{2}
    +\frac{k\alpha_2}{a^2\alpha_3}\frac{dc_g^2}{dt}
    \label{noncons3}.
\end{equation}
The first and last terms in the RHS are the same as in (\ref{non-cons}), for fixed $G_P$. The second term is new, and is not predicted by~\cite{AM}. We also have the extra 
Hamilton equations:
\begin{eqnarray}
    \dot T_R&=&-\frac{\partial H}{\partial \alpha_2}=\frac{\alpha_3}{\alpha_2}Na^3 \frac{\rho-3p}{2}\label{Ham1}\\
    \dot T_N&=&-\frac{\partial H}{\partial \alpha_3}=-Na^3\rho \label{Ham2}\\
    \dot \alpha_2&=&\frac{\partial H}{\partial T_R}=-Na \alpha_2 k \frac{\partial c_g^2}{\partial T_R}\label{Ham3}\\
    \dot \alpha_3&=&\frac{\partial H}{\partial T_N}=-Na \alpha_2 k \frac{\partial c_g^2}{\partial T_N},\label{Ham4}
\end{eqnarray}
where, we stress, one should keep $A^2$ (and not $a^2$) fixed evaluating the derivatives. This applies to the $a$ appearing in the matter Hamiltonian, and it makes a crucial difference for the $\dot T_R$ formula. 
The first formula agrees with (\ref{TRicci}) reduced to MSS. The second agrees with (\ref{TMatter}), modulo the issues discussed
in Section~\ref{LLIVmatter} (if we do not perform a LLIV integration by parts, $\dot T_N=Na^3p$, or pressure time). 
In the last two equations we have assumed that only $c_g$ enters $\bm \beta$, but these can (and will) be generalized.

Since we can replace:
\begin{equation}
    \frac{d c_g^2}{dt }= \sum_i \frac{\partial c_g^2}{\partial T_i} \dot T_i
\end{equation}
in the last term of (\ref{noncons3}), we find that the terms due to the dependence of $c_g$ on $T_R$ and $T_N$ cancel out the first two terms. That is, the dependence of $c_g$ on Ricci or Newtonian time does not lead to violations of energy conservation, with:
\begin{equation}
 \dot \rho+3\frac{\dot a}{a}(\rho+p)=0
\end{equation}
if $c_g$ only depends on $T_R$ and $T_N$ (if there are other dependences, these still contribute the usual terms).

This result it makes sense. 
A gravitational parameter depending on a clock leads to energy violations absorbed by that clock, but if the clock is gravitational there is nothing ``material'' to absorb the violations of energy conservation. So, there are none. In the language of natural selection, the  evolution is sterile.


\subsection{More general theories with varying Planck scale and gravitational coupling}\label{alpha2-other}

There is a final twist in the pattern: a {\it matter} parameter depending on a gravitational clock. The cancellations in the previous subsection do not apply if, for example, $c_m(T_R,T_N)$. Then, Eq.~(\ref{noncons3}) still holds true (since it only depends on the Einstein/gravity equations), but (\ref{Ham3}) and (\ref{Ham4}) receive a new term:
\begin{eqnarray}
     \dot \alpha_2&=&\frac{\partial H}{\partial T_R}=-Na \alpha_2 k \frac{\partial c_g^2}{\partial T_R}+Na^3\alpha_3\frac{\partial \rho}{\partial c_m^2}\frac{\partial c_m^2}{\partial T_R}\nn\\
         \dot \alpha_3&=&\frac{\partial H}{\partial T_N}=-Na \alpha_2 k \frac{\partial c_g^2}{\partial T_N}+Na^3\alpha_3\frac{\partial \rho}{\partial c_m^2}\frac{\partial c_m^2}{\partial T_N},\nn
\end{eqnarray}
resulting in:
\begin{equation}
 \dot \rho+3\frac{\dot a}{a}(\rho+p)=N a^3 \frac{\partial\rho}{\partial c_m^2}\left(\frac{\alpha_3}{\alpha_2}\frac{\partial c_m^2}{\partial T_R}\frac{\rho -3p }{2}-
 \frac{\partial c_m^2}{\partial T_N}\rho\right).\nn
\end{equation}
Had we considered multiple components, we would find that the general pattern is that if a matter $\bm\beta$ depends on a gravitational clock, then there are violations of energy conservation, apportioned between the different components by the dependence of their energy density on the relevant $\bm\beta$.


\section{General formula}
We now collect the various example we have studied in a single general formula. We should stress that 
the cancellations in Section~\ref{alpha2} are not  
mysterious. As we will show in a sequel to this paper~\cite{sequel}, they arise from the  sympletic Hamiltonian structure of the theory 
alone. 
The same applies to the general formula 
we will derive in this Section. It results from that structure plus the secondary 
constraint:
\begin{equation}
     \dot H\approx 0
\end{equation}
following from $H\approx 0$. This constraint
usually amounts to local matter/energy density conservation and the Bianchi identities for the geometry, whereas here it will lead to a generalized non-conservation formula.

\subsection{General formula in Einstein-Cartan theory}\label{genform1}
We first derive the general formula with reference to the Einstein-Cartan action subject to LLIV as defined in Section~\ref{LLIV} (and the assumption of homogeneity and isotropy). The idea is to 
repeat the calculations in Section~\ref{formaldevs} for a general ${\bm \beta}(T_{\bm\alpha})$. Defining:
\begin{equation}\label{Hsplit}
    H=H_G+H_M=\alpha_2 H_g+\alpha_3 H_m=
    \alpha_2 H_g+ \alpha_3Na^3\rho\nn
\end{equation}
so that $H_g$ does not depend on $\alpha_2$ (or $\alpha_3)$,
and evaluating $\dot H=0$ we find\footnote{Note that we need to use Hamilton's equations to eliminate $b$, and this contains new terms in $\dot \alpha_2$; hence the need to separate them. }:
\begin{eqnarray}
    \dot\rho+3\frac{\dot a}{a}(\rho+p)&=&-
    \frac{\dot \alpha_3}{\alpha_3}\rho +\frac{\dot\alpha_2}{\alpha_2}\frac{\rho-3p}{2}\nn\\
    &&-\frac{\alpha_2}{Na^3\alpha_3}\left(\sum_I\frac{\partial H_g}{\partial\alpha_I}\dot \alpha_I+
    \sum_{K}\frac{\partial H_g}{\partial\beta_K}\dot \beta_K
    \right).\nn
\end{eqnarray}
Using the Poisson equations,
the term in brackets can be written as:
\begin{eqnarray}
 \sum_{IK}
     \frac{\partial H_g}{\partial \alpha_I}\frac{\partial \beta_K }{\partial T_I}\frac{\partial H}{\partial \beta_K}-
\frac{\partial H_g}{\partial \beta_K}\frac{\partial \beta_K }{\partial T_I}\frac{\partial H}{\partial \alpha_I},
  \nn
\end{eqnarray}
and for $I=2,3$ rearranged as:
\begin{eqnarray}
&&-\frac{1}{\alpha_2}  \sum_{\substack{I=2,3\\K}}
\frac{\partial}{\partial \beta_K}(H-\alpha_3N a^3\rho)\frac{\partial \beta_K }{\partial T_I}\frac{\partial H}{\partial \alpha_I}
  \nn\\
&=&  -\frac{1}{\alpha_2}\left(\dot\alpha_2\dot T_2+\dot\alpha_3\dot T_3-\alpha_3N a^3\sum_{\substack{I=2,3\\K}}
\frac{\partial\rho }{\partial \beta_K}\frac{\partial \beta_K }{\partial T_I}\frac{\partial H}{\partial \alpha_I}
\right)\nn
\end{eqnarray}
so that the first two terms generally cancel the new terms in the continuity equation. For $I\neq 2,3$ only the matter component in $H$ leads to a non-zero contribution.  Thus:
\begin{eqnarray}
    \dot\rho+3\frac{\dot a}{a}(\rho+p)&=&\alpha_2 \sum_{\substack{I\neq 2,3\\K}}
     \frac{\partial\rho}{\partial \alpha_I}\frac{\partial \beta_K }{\partial T_I}\frac{\partial H_g}{\partial \beta_K}-
\frac{\partial\rho}{\partial \beta_K}\frac{\partial \beta_K }{\partial T_I}\frac{\partial H_g}{\partial \alpha_I}
    \nn\\
    &&-\sum_{\substack{I=2,3\\K}}
\frac{\partial\rho }{\partial \beta_K}\frac{\partial \beta_K }{\partial T_I}\frac{\partial H}{\partial \alpha_I}
\end{eqnarray}
with the last term being nothing but a generalization of the terms discussed in Section~\ref{alpha2-other}. In fact, since $\rho$ does not depend on $\alpha_2$ or $\alpha_3$, and $\frac{\partial H}{\partial\alpha_2}=
\frac{\partial H_G}{\partial\alpha_2}=H_g
$, this can be compressed into:
\begin{eqnarray}\label{GenF1}
    \dot\rho+3\frac{\dot a}{a}(\rho+p)&=&\sum_{\substack{I K}}
     \frac{\partial\rho}{\partial \alpha_I}\frac{\partial \beta_K }{\partial T_I}\frac{\partial H_G}{\partial \beta_K}-
\frac{\partial\rho}{\partial \beta_K}\frac{\partial \beta_K }{\partial T_I}\frac{\partial H_G}{\partial \alpha_I}
    \nn\\
    &&-Na^3\rho \sum_{\substack{K}}
\frac{\partial\rho }{\partial \beta_K}\frac{\partial \beta_K }{\partial T_N},
\end{eqnarray}
the last term being the only one that does not fit the simple pattern. 

\subsection{Generalization}\label{genform2}
 But this result is general and does not even require an action formulation: the sympletic/Hamiltonian structure is enough~\cite{sequel}.
An abridged proof 
is provided here (with a more general proof presented elsewhere), 
by ignoring the details of gravity and deriving the same result directly from the matter degrees of freedom. 

We assume a set of perfect fluids under homogeneity but that is not needed. The crucial inputs are the dependence of each specie energy density $\rho_i$ on $\alpha_I$ and $\beta_K$:
\begin{equation}
    \rho_i=\rho_i\left(\alpha_I, \beta_K\right)=\rho_i\left(\frac{\Pi_i}{a^3},\alpha_j,\beta_K \right).
\end{equation}
The diagonal dependence on $\Pi_i$ is non-negotiable, so 
we separated it and gave it a different index ($i$) from the other $\alpha_j$ (within a global $I$ index for $\bm\alpha$).
The other dependences are specific to each theory (see Section~\ref{cmingamma} for examples). For $\alpha_3$ we 
assume an implementation where $\rho_i$ does not depend on it; instead the $\Theta_i$ or $T_{m i}$ absorb it in their definition (cf. Eqs.~(\ref{SMcm2}) and~(\ref{SMcm1})). 

Then:
\begin{eqnarray}
    \dot\rho_i&=&\frac{\partial\rho_i}{\partial n_i}\dot n_i+\sum _{j} \frac{\partial\rho_i}{\partial \alpha_j}\dot \alpha_j
   + \sum _{K} \frac{\partial\rho_i}{\partial \beta_K}\dot \beta_K
    \nn
\end{eqnarray}
with 
\begin{eqnarray}
  \frac{\partial\rho_i}{\partial n_i}\dot n_i  &=&(p_i+\rho_i)\left(\frac{\dot\Pi_i}{\Pi_i}-3\frac{\dot a}{a}\right)\nn\\
  \frac{\partial\rho_i}{\partial \alpha_j}\dot\alpha_j
&=&    \frac{\partial\rho_i}{\partial \alpha_j}\frac{\partial H}{\partial T_j}
=\sum_K \frac{\partial\rho_i}{\partial \alpha_j}\frac{\partial H}{\partial \beta_K}\frac{\partial \beta_K}{\partial T_j}
\nn\\
\frac{\partial\rho_i}{\partial \beta_K}\dot \beta_K
&=&-\sum _I\frac{\partial\rho_i}{\partial \beta_K}
\frac{\partial \beta_K}{\partial T_I}
\frac{\partial H}{\partial \alpha_I }
\end{eqnarray}
where we have used (\ref{pdef}) and $\Pi_i=n_i a^3$ in the first equation, and Hamilton's equations for $\dot \alpha_j$ and $\dot T_I$ in the others. But the Hamilton equation for $\Pi_i$ implies\footnote{Or an equivalent expression for $m_i$, had we used it instead of $\Pi_i$; canonical transformations do not affect this formula.}:
\begin{equation}
    (p_i+\rho_i)\frac{\dot\Pi_i}{\Pi_i}=
    \frac{\partial\rho_i}{\partial \Pi_i}\frac{\partial H}{\partial T_{\pi i}}
\end{equation}
so this term fits into the same pattern as other $\bm\alpha$, once the metric dependence (the term in $\dot a/a$) has been separated. Hence:
\begin{equation}
     \dot\rho_i+3\frac{\dot a}{a}(p_i+\rho_i)=
     \sum_{I K}
     \frac{\partial\rho_i}{\partial \alpha_I}\frac{\partial \beta_K }{\partial T_I}\frac{\partial H}{\partial \beta_K}-
\frac{\partial\rho_i}{\partial \beta_K}\frac{\partial \beta_K }{\partial T_I}\frac{\partial H}{\partial \alpha_I}.
\end{equation}
This is the most informative equation we have. It tells us, specie by specie, the violations of energy conservation for arbitrary $\bm\beta$ depending on arbitrary times $\bm T_{\bm\alpha}$. 

Summing over $i$ we finally get:
\begin{eqnarray}\label{GenF2}
     \dot\rho+3\frac{\dot a}{a}(p+\rho)&=&
      \sum_{I K}
     \frac{\partial\rho}{\partial \alpha_I}\frac{\partial \beta_K }{\partial T_I}\frac{\partial H}{\partial \beta_K}-
\frac{\partial\rho}{\partial \beta_K}\frac{\partial \beta_K }{\partial T_I}\frac{\partial H}{\partial \alpha_I}\nn\\
&=&
      \sum_{I K}
     \frac{\partial\rho}{\partial \alpha_I}\frac{\partial \beta_K }{\partial T_I}\frac{\partial H_G}{\partial \beta_K}-
\frac{\partial\rho}{\partial \beta_K}\frac{\partial \beta_K }{\partial T_I}\frac{\partial H_G}{\partial \alpha_I}\nn\\
&&-Na^3\rho \sum_{\substack{K}}
\frac{\partial\rho }{\partial \beta_K}\frac{\partial \beta_K }{\partial T_N},
\end{eqnarray}
that is, Eq.~\ref{GenF1}, derived without reference to the gravity action, and so independent of the gravitational dynamics.

The above can be written covariantly with  replacements $a^3\rightarrow \sqrt{h}$ and $3 \dot a/a\rightarrow\theta= \dot h/h$.
Defining
$   H_m=N\sqrt{h}\rho=N\sqrt{h} \sum _i\rho_i$
the general covariant expression for non-conservation is:
\begin{widetext}
\be\label{GenEq}
\boxed{
     n^\nu\nabla_\mu T^\mu_{\;\; \nu}=
     \frac{1}{N\sqrt{h}\alpha_3}
     \sum_{I K}
\frac{\partial \beta_K }{\partial T_I}\left(
     \frac{\partial H_m}{\partial \alpha_I}\frac{\partial H_G}{\partial \beta_K}-
\frac{\partial H_m}{\partial \beta_K}
\frac{\partial H_G}{\partial \alpha_I}\right)
-N\sqrt{h}\rho \sum_{\substack{K}}
\frac{\partial\rho }{\partial \beta_K}\frac{\partial \beta_K }{\partial T_N}.}
\end{equation}
\end{widetext}
where $n^\mu$ is the normalized normal to $\Sigma$. Modulo $\alpha_3$ issues the right hand side is just the non-local component of the Poisson bracket $\{H_m,H_g\}$ (as explored further in~\cite{sequel}).

\subsection{General pattern}
In view of this general formula
we can collate and confirm the various patterns we found
in the previous Sections.

As its first term indicates, to get net non-conservation of energy one possibility is for a gravitational parameter (a $\bm\beta$ appearing in $H_G$) to depend on one or more matter clocks (i.e. clocks dual to $\bm\alpha$ appearing in at least one matter variable). This is the case in Sections~\ref{firstmodels}, where $c_g$ depends on matter clocks ($T_\Lambda$ and $T_c$). The more a specie $i$ contributes to the dual constant of these clocks (according to a measure which is like a chemical potential for that constant), the more of a share of violations it gets. The gravitational parameter can also depend jointly on many clocks (e.g. via functions of the form $c_g(T_\Lambda,T_c)$), in which case the partial derivatives with respect to the clock time controls the apportionment. 

Its second and third terms open up another possibility for a net energy conservation violation: a matter parameter (a $\bm\beta$ in $H_M$) dependent on the time ticked by a gravity clock (dual to a $\alpha$ appearing in $H_G$, or to the prefactor $\alpha_3$ of $H_m$, describing the strength of the gravitational coupling). 
This is the case in Section~\ref{alpha2-other}, where $c_m=c_m(T_R,T_N)$. The violations are shared by the different species according to how much their energy density depends on this matter parameter. 

No other combination generates net energy. A purely matter parameter dependent on a purely matter clock is similar to an interaction term, energy exchanging between different components, but with no net gain. This is the case in Section~\ref{cmingamma}, with $c_m(T_\Lambda)$ if $c_g$ is left fixed (or no other gravity parameter depended on $T_\Lambda$). As we will see in Section~\ref{Lambda-rems}, this is not useless: for example Lambda could exchange energy with other forms of matter via this mechanism, leading to an interesting cosmogony and opportunities to ``solve'' the $\Lambda$ problem.

Only a purely gravity parameter depending on a gravity clock would lead to no energy conservation violations in {\it any} matter component. This is the case explored in Section~\ref{alpha2}, with $c_g=c_g(T_R,T_N)$ (and $\alpha_3$ treated so that it does not appear in $H_m$). Such situation is entirely sterile from the point of view of matter production.


\section{Illustrative LLIV cosmogonies}
\label{higgle}
We finally provide some examples of scenarios for the origin of the matter in the Universe, focusing on LLIV (deferring non-LLIV scenarios to~\cite{sequel}). We can think of the potentials $\bm\beta(\bm\alpha)$ as random choices for chaotic evolution, selected in/out by their ultimate effects, specifically by whether they lead to ``viable'' offspring. 

\subsection{VSL scenarios}
Foremost we find VSL scenarios relating matter production to departures from spatial flatness. For example we could take  $c^2_m\in \bm \alpha$, $c^2_g\in\bm \beta$ (with $c_g(T_c)$) as in Section~\ref{cmingamma}, and  investigate a phase transition with:
\begin{equation}\label{step}
    c_g^2=c^2_{g-}H(T_{c\star}-T_c)+ c^2_{g+}H(T_c-T_{c\star}) 
\end{equation}
where $H$ is the Heaviside step function and $T_{\star}$ is a critical time (possibly, but not necessarily of the order of the Planck scale). From the Hamilton equations for $a$ and $b$ we know that for such a shock there can only be delta functions in $\dddot a$, so $a$ and $\dot a$ must be continuous. From the Hamiltonian constraint, 
$a_-=a_+=a_\star$ and $\dot a_-=\dot a_+=\dot a_\star$ implies:
\begin{equation}\label{Deltarho}
    \Delta\rho=\frac{k\alpha_2}{a_\star^2\alpha_3 }\Delta c_g^2,
\end{equation}
which could also be derived directly from (\ref{non-cons}). In such a scenario:
\begin{eqnarray}
    a&=&c_{g-}\sqrt{|k|} t \qquad \qquad\qquad(t<t_\star)\nn\\
    &=&c_{g-}\sqrt{|k|} \left(\frac{t}{t_\star}
\right)^{1/2}\qquad (t>t_\star)
\end{eqnarray}
the only discontinuity happening in $\ddot a$. 

This much just mimics the calculations in~\cite{AM}, but we can say more. The jump in $c_m^2$ can be computed as:
\begin{equation}\label{cmcg}
    \frac{d c_m^2}{dt}=\frac{\partial H}{\partial T_c}=-\frac{Na\alpha_2 k}{\dot T_c} \frac{d c_g^2}{dt}=\frac{k\alpha_2 }{a^2\alpha_3\mu_c} \frac{d c_g^2}{dt}.
\end{equation}
This is a non-linear equation (since $\mu_c=\mu_c(c_m^2)$) with solution:
\begin{equation*}
    \int _{c_{m-}^2}^ {c_{m+ }^2}      \mu_{ c} \, d c_m^2=\frac{k\alpha_2 }{a_\star ^2\alpha_3} \Delta c_g^2
\end{equation*}
so that indeed:
\begin{equation}\label{deltarhocmcg}
   \Delta\rho=\rho(c_{m+}^2)-\rho (c_{m-}^2)=\frac{k\alpha_2}{a_\star^2\alpha_3 }\Delta c_g^2  
\end{equation}
i.e. Eq.~(\ref{Deltarho}). But we also learn that the violations are shared according to:
\begin{equation}\label{deltarhoicmcg}
         \Delta\rho_i=\rho_i(m_i,c_{m+}^2)-\rho_i(m_i,c_{m-}^2)=\int _{c_{m-}^2}^ {c_{m+ }^2}      \mu_{i c} \, d c_m^2.
\end{equation}
Notice that at least one $m_i$ has to be non-zero, otherwise $\dot T_c=0$ (see~Eq.(\ref{Tcm})) and $c_g$ would never feel the step predicted in  $c_g^2=c_g^2(T_c)$. More generally $\mu_c\propto \rho$, so the less matter there is to begin with, the bigger the jump in $c_m$. 
In this theory the net violations are  (\ref{Deltarho}), since   this only depends on the gravitational action/equations, but in addition $c_m$ must change so that this $\Delta\rho$ is produced given the overall dependence $\rho(c_m^2)$, with the various species receiving a contribution in proportion to their individual dependence $\rho_i(c_m^2)$. We could have played this game with any other fundamental matter parameter and its clock, say $e$, $m_e$, etc. 

We could instead have used the reverse situations, with  $c^2_g\in \bm \alpha$ and $c^2_m\in\bm \beta$. The gravitational clock associated with $c_g^2$ ticks as
\begin{equation}
    \dot T_{cg}=\frac{\partial H}{\partial c_g^2}=Na\alpha_2 k,
\end{equation}
and we could consider a step function $c_m^2=c_m^2(T_{cg})$. 
As it happens there is a formal symmetry between $c_m^2=c_m^2(T_{cg})$ and 
$c_g^2=c_g^2(T_c)$, because:
\begin{equation}
    \frac{dc_m^2}{dc_g^2}=\frac{\frac{dc_m^2}{dt}}{\frac{dc_g^2}{dt}}=-\frac{\frac{\partial H}{\partial c_g^2}}{\frac{\partial H}{\partial c_m^2}}=\frac{k \alpha_2}{a^2\alpha_3\mu }
\end{equation}
i.e. $d\beta/d\alpha$ drops out of the calculation leading to  (\ref{cmcg}), whatever choice we make for $\alpha$ and $\beta$. We then have the same end results (\ref{deltarhocmcg}) and (\ref{deltarhoicmcg}). 

The physics is very different, however. Whereas for $c_g^2(T_c)$ we need some matter for the step function to be felt ($T_c\propto \rho$), here we need $k\neq 0$, i.e. some spatial curvature. Otherwise $T_{cg}$ would stop and $c_m$ never feel the jump. Hence, 
the $T_{cg}$ interpretations sheds light on
the flatness problem. Such a curvature clock stops if $k=0$ or if curvature becomes negligible (just as a Ricci clock stops for radiation). 
It also flows in opposite directions for $k=\pm 1$. Flatness therefore becomes an attractor, with 
$\Delta c_m^2/\Delta c_g^2$ having the sign of $k/\mu$. 

\subsection{Lambda remittances}
\label{Lambda-rems}
In the above model the $\Lambda$ problem is simply the statement that $\mu_{\Lambda c}$ is much smaller than the other $\mu_{i c}$ if $\Lambda=0$ originally (with further fine tuning needed to address the quasi-Lambda problem~\cite{quasi-Barrow}). Whilst it is possible that this can be proved from first principles, we should also explore models 
allowing the vacuum energy to change as part of the evolution itself, dumping its energy into a form of matter. Such {\it vacuum remittances}
can be achieved either with $\Lambda$ the generator of a clock ($\rho_\Lambda\in\bm \alpha$), with  a matter $\bm\beta$ depend on $T_\Lambda$ (e.g. $c_m=c_m(T_\Lambda)$); or with $\rho_\Lambda\in\bm\beta$ and dependent on a matter clock,  such as with $\rho_\Lambda(T_c)$.

For example, with $c_m=c_m(T_\Lambda)$, we can adapt the calculations in Section~\ref{cmingamma} to find:
\begin{eqnarray}
     \dot \rho_i+3\frac{\dot a}{a}(\rho_i+p_i)
     &=&
     \frac{m_i\dot f_i}{a^{3(1+w_i)}}=
\frac{m_i f'_i}{a^{3(1+w_i)}}
     \frac{dc_m^2}{dt};\; (i\neq \Lambda) 
     \nn  \\
        \dot \rho_\Lambda
     &=&
     -\sum_{i\neq \Lambda} 
     \frac{m_i f'_i}{a^{3(1+w_i)}}
     \frac{dc_m^2}{dt},\label{remittcmTL}
\end{eqnarray}
so that for a shock of the form (\ref{step}):
\begin{eqnarray}
     \Delta\rho_\Lambda&=&-\sum_{i\neq \Lambda}\Delta\rho_i \nn\\
     \Delta\rho_i&=&\rho_i(c_{m+}^2)-\rho_i(c_{m-}^2)=\int _{c_{m-}^2}^ {c_{m+ }^2}      \mu_{i c} \, d c_m^2;\; (i\neq \Lambda) .  \nn
\end{eqnarray}
As expected, 
if we include $\Lambda$ in the accounts, then no net energy is produced; however there is energy exchange, with  a $\Lambda$ remittance into the rest of the matter, its 
energy distributed by species according to their $\mu_{ic}$. This could be done with any other matter parameter elevated to $\bm\beta$. 

Alternatively, we could have $\rho_\Lambda=m_\Lambda f_\Lambda(c_m^2)$ and $m_\Lambda\in\bm\beta$ and $c_m^2\in\bm\alpha$ (i.e. a $m_\Lambda(T_c)$ potential).  Then:
\begin{eqnarray}
    \frac{dc_m^2}{dt}=\frac{\partial H}{\partial T_c}&=&Na^3\frac{\partial\rho_\Lambda}{\partial T_c}= Na^3f_\Lambda m'_\Lambda
\end{eqnarray}
so that instead of (\ref{dotrhocgtc}) we find
for $i\neq \Lambda$:
\begin{eqnarray}
     \dot \rho_i+3\frac{\dot a}{a}(\rho_i+p_i)
     &=&
\frac{m_i f'_i}{a^{3(1+w_i)}}
     \frac{dc_m^2}{dt}=\frac{Nm_i f'_i}{a^{3w_i}}f_\Lambda m'_\Lambda\nn
\end{eqnarray}
which, summed over $i$, leads to $-\dot\rho_\Lambda=-(\dot m_\Lambda f_\Lambda +m_\Lambda \dot f_\Lambda)$ (recall that $\Lambda$ may contribute to (\ref{Tcm})).

In both scenarios, whatever caused the hot Big Bang suppressed the vacuum energy. The Big Bang is a sign of variability involving Lambda and a matter parameter, with one of them playing the role of clock. 

\subsection{Latter-day Big Bangs?}\label{latterBB}
Naturally within this framework one may ask why would the Big Bang be unique? And could this be related to why Lambda is so small compared to the Planck scale? The simplest scenario in this respect is a cascade of Big Bangs  siphoning matter out of Lambda when Lambda resurfaces. A Big Bang in the past, originated by a phase transition due to vacuum domination, would be reflected in a Big Bang in the future when similar circumstances arise (as seems to be the case around now). 

However, in such a cyclic scenario, the Big Bang temperature would be $10^{-32}$ smaller each time, which is nonsense. We should therefore renormalize the Planck mass at the start of each cycle. This could happen in two ways: either via a dependence on Ricci time, $T_R$, or by making $\alpha_2$ a $\bm \beta$. 
Imagine a $\rho_\Lambda(T_c,T_R)$. 
Then
\begin{equation}
    \dot\alpha_2=Na^3\frac{\partial \rho_\Lambda}{\partial T_R}
\end{equation}
and beside the remittance (\ref{remittcmTL}), we have the correction:
\begin{eqnarray}
      \dot\rho_\Lambda&=&
        -\sum_{i\neq \Lambda} 
     \frac{m_i f'_i}{a^{3(1+w_i)}}
     \frac{dc_m^2}{dt}
      +2
      \frac{Na^3 }{\alpha_2}\rho_\Lambda\frac{\partial \rho_\Lambda}{\partial T_R}.
\end{eqnarray}
We can design $\rho_\Lambda(T_c,T_R)$ so that the Planck scale changes at each event and any new Big Bang event is at the Planck scale. 


\section{Conclusions and outlook}
Where does all the matter in the Universe come from? In this paper we proposed that the Universe acquired its energy by virtue of ``changing the rules of the game'', viz. evolution in the laws of physics. Rather than implementing evolution in terms of a coordinate time, we defined it in terms of time variables, $\bm T_{\bm\alpha}$, dual to constants, $\bm \alpha$, using as a blueprint one formulation of unimodular gravity~\cite{unimod} (in which $\Lambda$ plays the role of $\bm\alpha$). Variability was then introduced by making {\it other} constants, $\bm\beta$, functions of these times, $\bm\beta(\bm T_{\bm\alpha})$. The main practical advantage over coordinate time dependence, is that this approach specifies where the energy is credited. 
In general there is energy production if either a matter parameter varies in terms of a gravitational clock, or a gravity parameter evolves in terms of a matter clock, with that matter component acquiring the energy violation balance. Other  combinations are sterile.


As in Wheeler's view, we can think of the evolution potentials $\bm\beta(\bm T_{\bm\alpha})$ as random choices for chaotic evolution, selected in or out by their ultimate effects. 
These could be 
whether ``viable'' offspring is produced: only evolution that implies creation of matter is beneficial, with sterile evolution edited out. In this paper we ``geocentrically'' qualified this statement as creation of {\it semi-classical} matter, but this is not strictly necessary. Perhaps the quantum cosmology of this process is far more interesting, with classical matter creation replaced by an $S$-matrix process. In Appendix~\ref{QC} we sketch the first steps in this direction, to show that progress is possible, but defer full treatment to future work. Another offshoot is a classical realization of the Hartle-Hawking initial conditions, as sketched in Appendix~\ref{signature}. This leads to signature change, a possibly  
essential feature for more detailed model building, since it affects the horizon structure of the Universe (see Appendix~\ref{signature}).


We stress that the proposal in this paper is distinct from multidimensional time. On-shell, the various ${\bm T}_{\bm\alpha}$ are all a function of each other, so time is classically one dimensional. But the ${\bm T}_{\bm\alpha}$ start off as independent variables (something which has dramatic effects in the quantum theory~\cite{gielen,gielen1,JoaoLetter,JoaoPaper}), and index evolution differently should there be a $\bm\beta(\bm T_{\bm\alpha})$. If there are violations of energy conservation due to evolution, it is crucial to specify in terms of which of these clocks evolution takes place (even though on-shell they are all related). This has physical effects, namely it determines into which matter component the energy goes/is taken off. 

We close with a few comments on future developments. 
In this paper we focused on theories exhibiting local Lorentz invariance violations, defined in the spirit of Horava-Lifshitz theory: a $4=3+1$ split that leads to a $3+1\neq 4$ after one fiddles with  some aspect of the split theory. This author started off this article with the hunch that Lorentz invariance violation would be favoured by cosmic procreation of matter, but was disabuse by hard calculations. The shattered remains of this dogma lead to the sequel~\cite{sequel}: a repeat of this paper without recourse to local Lorentz invariance violation.  We should stress that in our theories, even under LLIV and with time-dependence, we have a sympletic/Hamiltonian structure. This is very powerful, and is responsible for most of the results in this paper.

Finally, there is the question of falsifiability. 
The various scenarios with matter progeny considered in Section~\ref{higgle} are all equally possible, so one may raise questions about testability. But producing copious amounts on matter on near homogeneous patches is just the starting point: one must add fluctuations. We gave special attention to scenarios with bimetric VSL ($c_m\neq c_g$) because these are known to produce very distinctive predictions for the cosmic fluctuations~\cite{nia,nia1} (see also~\cite{Moffatbim,cssound,bim1,bim2}). More directly, we speculate on whether violations of energy conservation might be accessible on not-so-large scales, for example during black hole formation. This is particularly relevant if there is evolution in the laws of physics in our future, as predicted Section~\ref{latterBB}. In the presence of foliation-dependent evolution, the region near a black hole horizon acts as a crystal ball for events in our future. Any future or latter-day Big Bangs that might be implied, will indeed happen near the black hole horizon, with possible observational consequences.  This matter is currently under investigation.

\section{Acknowledgments}
We thank Bruno Alexandre, Michele Arzano, Giulia Gubitosi, John Moffat, Alex Vikman and an anonymous referee for shaping this paper. This work was partly supported by the STFC Consolidated Grants ST/T000791/1 and
ST/X00575/1.

\appendix
\section{Quantum cosmology}
\label{QC}
We briefly sketch how a quantum treatment might start. 
Standard theories with relational times $\bm T_{\bm\alpha}$ are known to convert the Wheeler-DeWitt equation into a Schr\"odinger equation (e.g.~\cite{gielen,gielen1,JoaoLetter,JoaoPaper}).  An independent claim to the same effect was made for VSL scenarios with hard break of diffeomorphism invariance~\cite{MoffatQC}.
For the VSL scenarios we have proposed in this paper we obtain a time-dependent Schr\"odinger equation, similar to what one gets in the interaction picture. Hence the classical matter creation we have reported can also be seen as quantum matter creation, and an $S$-matrix approach could be developed, as we now show.

To fix ideas, we take unimodular VSL (see Section~\ref{unimodVSL}). One can put the Hamiltonian constraint in the form~\cite{JoaoPaper}:
\be
0=\frac{1}{b^2+kc_g^2}a^2-\frac{3}{\Lambda}\equiv 
H_0 -\phi\nn
\ee
with $\Lambda$ replaced by $\phi=3/\Lambda$ and the dynamical Hamiltonian $H_0$ defined by this expression. Choosing $\phi$ (instead of $\Lambda$) for $\alpha$ and allowing $c_g^2$ to become a function of $T_\phi$, upon quantization with suitable ordering we obtain the time-dependent Schr\"odinger equation:
\be\label{WDWSchro1}
\left[\hat H_0 (b,T_\phi) - i \plk \frac{\partial }{\partial T_\phi}\right]\psi(b,T_\phi) =0,\nn
\ee
with 
\be
\hat H_0(b,T_\phi)= 
\frac{-i\plk}{b^2+kc_g^2(T_\phi)} \frac{\partial}{\partial b}\nn
\ee
following from (\ref{commutator}) (where $\plk$ is defined). As is well known, if $c_g$ does not change, one possible solution is:
\be
\psi(b,T_\phi;\phi)={\cal N}e^{-i\frac{\phi}{\plk}T_\phi}\psi_s(b,\phi)\nn
\ee
where 
\begin{equation}
    \psi_s(b,\phi)=e^{i\frac{\phi}{\plk}X_{CS}(b)}=
    e^{i\frac{\phi}{\plk}\left(\frac{b^3}{3}+kc_g^2b\right)}\nn
\end{equation}
is a minisuperspace version of the Chern-Simons-Kodama state~\cite{kodama,realkod,stephkod}. 
This is the monochromatic solution and superpositions with different $\Lambda$ are possible (see~\cite{JoaoPaper,stephkod} for a discussion of normalizability and inner product). If $c_g^2$ changes, 
the solution is instead a time-dependent version of the Chern-Simons-Kodama  state:
\begin{equation}
    \psi(b,T_\phi)={\cal N} {\cal T} \exp{\left[-\frac{i}{\plk }\int^{T_{\phi}}_{T_{\phi 0}} H_0(\tilde T_\phi) d\tilde T_\phi \right]}\psi(b,T_{\phi 0})\nn
\end{equation}
where $\cal T$ denotes time-ordering in $T_\phi$, thus opening up a parallel with the $S$ matrix framework. 

\section{Signature change and evolution}\label{signature}
The fact that the natural variable in the dynamics is often $c^2$ suggests that
$c$  could appear in $\bm\alpha$ as a square. We could then entertain scenarios in which the sign of $c^2$, and so the signature, changes and this has a curious relation to the Hartle-Hawking~\cite{HH} no-boundary proposal, in a process reminiscent of~\cite{VilUni}. This could be done with:
\begin{equation}    c_g^2(T_\Lambda)=c_0^2H(T_\Lambda)- c_0^2 H (-T_\Lambda)
\end{equation}
for $k=1$ and $c_0^2>0$. Then, Eq.~(\ref{nonconsLambda}) implies:
\begin{equation}
    \Delta\rho_\Lambda=\frac{k\alpha_2 }{a_\star ^2\alpha_3} \Delta c_g^2=2c_0^2\frac{k\alpha_2 }{a_\star ^2\alpha_3} 
\end{equation}
which combined with the Hamiltonian constraint implies a swap in the sign of $\rho_\Lambda=\pm \rho_{\Lambda 0}$, equivalent to a redefinition of $N$ and:
\begin{equation*}
    \alpha_2(-b^2+kc_0^2)+\alpha_3\rho_{\Lambda 0}a^2\approx 0.
\end{equation*}
For $T_\Lambda<0$ we therefore obtain a solution to the (real) Euclidean theory: the 4-sphere. In this scenario, rather than changing signature as a function of $b$ (as in~\cite{VilUni}), the change happens as a function of $T_\Lambda$, that is, with proper time evolution. It is curious that Hartle and Hawking's ``creation out of nothing'', usually seen as an instanton and so a purely quantum process, could in fact be a classical phenomenon, within a theory with varying laws and classical signature change. 
This would dramatically change the usual discussion of the horizon problem, which would thus become disconnected from the flatness problem. Options in which the matter speed of light increases (rather than decreases, as in the usual solution to the horizon problem) could then be considered, as long as the sign of $c^2$ changed. We could also have early Universe situations where $c_m^2>0$ but $c_g^2<0$, that is Lorentzian spacetime for matter (or some of matter) and Euclidean gravity, or vice versa.


\begin{thebibliography}{99}

\bibitem{PaulDavis}
Paul Davis, https://metanexus.net/large-cosmic-lessons-physics/



\bibitem{Smolin}
L.~Smolin,
[arXiv:2202.00594 [gr-qc]]; 
[arXiv:2105.03539 [quant-ph]].

\bibitem{Nielsen}
S.~Chadha and H.~B.~Nielsen,
Nucl. Phys. B \textbf{217}, 125-144 (1983)
doi:10.1016/0550-3213(83)90081-0; D.~Forster, H.~B.~Nielsen and M.~Ninomiya,
Phys. Lett. B \textbf{94}, 135-140 (1980)
doi:10.1016/0370-2693(80)90842-4.

\bibitem{dirac}
P.A.M. Dirac, Nature (London) 139 (1937) 323.

\bibitem{anth}
J. Barrow and F. Tipler,
The Anthropic Cosmological Principle, Oxford Paperbacks, 1989. 

\bibitem{EllisVSL}
G.~F.~R.~Ellis and J.~P.~Uzan,
Am. J. Phys. \textbf{73}, 240-247 (2005)
doi:10.1119/1.1819929
[arXiv:gr-qc/0305099 [gr-qc]].

\bibitem{vikman}
P.~Jirou\v{s}ek, K.~Shimada, A.~Vikman and M.~Yamaguchi,
JCAP \textbf{04}, 028 (2021).

\bibitem{vikman1}
A.~Vikman,
[arXiv:2107.09601 [gr-qc]].


\bibitem{unimod} M.~Henneaux and C.~Teitelboim, ``The cosmological constant and general covariance,'' {\em Phys.\ Lett.\ B} {\bf 222} (1989), 195--199.

\bibitem{unimod1}
W. G. Unruh, Phys. Rev. {\bf D40}, 1048 (1989);  K.~V.~Kucha\v{r}, ``Does an unspecified cosmological constant solve the problem of time in quantum gravity?,'' {\em Phys.\ Rev.\ D} {\bf 43} (1991), 3332--3344.


\bibitem{UnimodLee1}
L.~Smolin, ``Quantization of unimodular gravity and the cosmological constant problems,'' {\em Phys.\ Rev.\ D} {\bf 80} (2009), 084003, arXive: 0904.4841. 

\bibitem{alan} A. Daughton, J. Louko, and R. D. Sorkin, ``Instantons and unitarity in quantum cosmology with fixed four-volume,'' {\em Phys.\ Rev.\ D} {\bf 58}, 084008 (1998).

\bibitem{daughton} A. Daughton, J. Louko, and R. D. Sorkin, ``Initial conditions and unitarity in unimodular quantum cosmology,'' [gr-qc/9305016].

\bibitem{sorkin1} R. D. Sorkin, ``Role of time in the sum-over-histories framework for gravity,'' {\em Int J Theor Phys} {\bf 33}, 523–534 (1994). https://doi.org/10.1007/BF00670514

\bibitem{sorkin2} R. D. Sorkin, ``Forks in the road, on the way to quantum gravity,'' {\em Int J Theor Phys} {\bf 36}, 2759–2781 (1997). https://doi.org/10.1007/BF02435709




\bibitem{Bombelli}
L.~Bombelli, W.~E.~Couch and R.~J.~Torrence,
Phys. Rev. D \textbf{44}, 2589-2592 (1991)
doi:10.1103/PhysRevD.44.2589

\bibitem{UnimodLee2}
L.~Smolin,
Phys. Rev. D \textbf{84}, 044047 (2011)
doi:10.1103/PhysRevD.84.044047
[arXiv:1008.1759 [hep-th]].





\bibitem{JoaoLetter}
J.~Magueijo,
Phys. Lett. B \textbf{820}, 136487 (2021)
doi:10.1016/j.physletb.2021.136487
[arXiv:2104.11529 [gr-qc]].

\bibitem{JoaoPaper}
J.~Magueijo,
Phys. Rev. D \textbf{106}, no.8, 084021 (2022)
doi:10.1103/PhysRevD.106.084021
[arXiv:2110.05920 [gr-qc]].


\bibitem{weinberg}
S.~Weinberg,
Rev. Mod. Phys. \textbf{61}, 1-23 (1989).

\bibitem{padilla}
A.~Padilla,
``Lectures on the Cosmological Constant Problem,''
[arXiv:1502.05296 [hep-th]].


\bibitem{pad}
N.~Kaloper and A.~Padilla,
Phys. Rev. Lett. \textbf{112}, no.9, 091304 (2014).

\bibitem{pad1}
N.~Kaloper, A.~Padilla, D.~Stefanyszyn and G.~Zahariade,
Phys. Rev. Lett. \textbf{116}, no.5, 051302 (2016)
doi:10.1103/PhysRevLett.116.051302
[arXiv:1505.01492 [hep-th]].

\bibitem{lomb}
L.~Lombriser,
Phys. Lett. B \textbf{797}, 134804 (2019).




\bibitem{AM}
A.~Albrecht and J.~Magueijo,
Phys. Rev. D \textbf{59}, 043516 (1999)
doi:10.1103/PhysRevD.59.043516
[arXiv:astro-ph/9811018 [astro-ph]].

\bibitem{MoffatVSL}
J.~W.~Moffat,
Int. J. Mod. Phys. D \textbf{2}, 351-366 (1993)
doi:10.1142/S0218271893000246


\bibitem{VSLreview}
J.~Magueijo,
Rept. Prog. Phys. \textbf{66}, 2025 (2003)
doi:10.1088/0034-4885/66/11/R04
[arXiv:astro-ph/0305457 [astro-ph]].


\bibitem{unimodProca}
R.~Isichei and J.~Magueijo,
[arXiv:2305.09380 [hep-th]].

\bibitem{vikaxion}
K.~Hammer, P.~Jirousek and A.~Vikman,
[arXiv:2001.03169 [gr-qc]].

\bibitem{Thiemann}
T. Thiemann, Modern Canonical Quantum General Relativity, CUP, Cambridge 2008. 

\bibitem{brown}
J.~D.~Brown,
Class. Quant. Grav. \textbf{10}, 1579-1606 (1993)
doi:10.1088/0264-9381/10/8/017
[arXiv:gr-qc/9304026 [gr-qc]].

\bibitem{gielen} 
S.~Gielen and L.~Men\'endez-Pidal,
Class. Quant. Grav. \textbf{37}, no.20, 205018 (2020).

\bibitem{gielen1}
S.~Gielen and L.~Men\'endez-Pidal,
Class. Quant. Grav. \textbf{39}, no.7, 075011 (2022)
doi:10.1088/1361-6382/ac504f
[arXiv:2109.02660 [gr-qc]].


\bibitem{twot}
B.~Alexandre and J.~Magueijo,
Phys. Rev. D \textbf{104}, no.12, 124069 (2021)
doi:10.1103/PhysRevD.104.124069
[arXiv:2110.10835 [gr-qc]].

\bibitem{vollick}
  D.~N.~Vollick,
  Phys.\ Rev.\ D {\bf 68}, 063510 (2003)
  doi:10.1103/PhysRevD.68.063510
  [astro-ph/0306630].

\bibitem{vollick2}
  D.~N.~Vollick,  
  Class.\ Quant.\ Grav.\  {\bf 21}, 3813 (2004)
  doi:10.1088/0264-9381/21/15/N01
  [gr-qc/0312041].

\bibitem{flanagan}
  E.~E.~Flanagan,
  Phys.\ Rev.\ Lett.\  {\bf 92}, 071101 (2004)
  doi:10.1103/PhysRevLett.92.071101


\bibitem{HL}
P.~Horava,
Phys. Rev. D \textbf{79}, 084008 (2009)
doi:10.1103/PhysRevD.79.084008
[arXiv:0901.3775 [hep-th]].

\bibitem{CovVSL}
J.~Magueijo,
Phys. Rev. D \textbf{62} (2000), 103521
doi:10.1103/PhysRevD.62.103521
[arXiv:gr-qc/0007036 [gr-qc]].



\bibitem{Afshordi}
N.~Afshordi and J.~Magueijo,
Phys. Rev. D \textbf{106}, no.12, 123518 (2022)
doi:10.1103/PhysRevD.106.123518
[arXiv:2209.07914 [hep-th]].

\bibitem{Lee2}
S.~Alexander, M.~Cort\^es, A.~R.~Liddle, J.~Magueijo, R.~Sims and L.~Smolin,
Phys. Rev. D \textbf{100}, no.8, 083507 (2019)
doi:10.1103/PhysRevD.100.083507
[arXiv:1905.10382 [gr-qc]].
\bibitem{tom}
J.~Magueijo and T.~Z\l{}o\'snik,
Phys. Rev. D \textbf{100}, no.8, 084036 (2019)
doi:10.1103/PhysRevD.100.084036
[arXiv:1908.05184 [gr-qc]].

\bibitem{Coleman}
S.~R.~Coleman and S.~L.~Glashow,
Phys. Rev. D \textbf{59}, 116008 (1999)
doi:10.1103/PhysRevD.59.116008
[arXiv:hep-ph/9812418 [hep-ph]].

\bibitem{Carroll}
S.~M.~Carroll, G.~B.~Field and R.~Jackiw,
Phys. Rev. D \textbf{41}, 1231 (1990)
doi:10.1103/PhysRevD.41.1231

\bibitem{MoffatQC}
J.~W.~Moffat,
Found. Phys. \textbf{23}, 411-437 (1993)
doi:10.1007/BF01883721
[arXiv:gr-qc/9209001 [gr-qc]].

\bibitem{sequel}
J. Magueijo, in preparation. 

\bibitem[Hartle and Hawking(1983)]{HH}
J.~B.~Hartle and S.~W.~Hawking,
``Wave Function of the Universe,'' {\em Phys.\ Rev.\ D} \textbf{28} (1983), 2960--2975.

\bibitem{VilUni}
B.~Alexandre, R.~Isichei and J.~Magueijo,
[arXiv:2304.00666 [hep-th]].



\bibitem{quasi-Barrow}
J.~D.~Barrow and J.~Magueijo,
Phys. Lett. B \textbf{447}, 246 (1999)
doi:10.1016/S0370-2693(99)00008-8
[arXiv:astro-ph/9811073 [astro-ph]].


\bibitem{Moffatbim}
M.~A.~Clayton and J.~W.~Moffat,
Phys. Lett. B \textbf{460}, 263-270 (1999)
doi:10.1016/S0370-2693(99)00774-1
[arXiv:astro-ph/9812481 [astro-ph]].

\bibitem{cssound}
J.~Magueijo,
Phys. Rev. Lett. \textbf{100}, 231302 (2008)
doi:10.1103/PhysRevLett.100.231302
[arXiv:0803.0859 [astro-ph]].

\bibitem{bim1}
J.~Magueijo,
Phys. Rev. D \textbf{79}, 043525 (2009)
doi:10.1103/PhysRevD.79.043525
[arXiv:0807.1689 [gr-qc]].

\bibitem{bim2}
J.~Magueijo, J.~Noller and F.~Piazza,
Phys. Rev. D \textbf{82}, 043521 (2010)
doi:10.1103/PhysRevD.82.043521
[arXiv:1006.3216 [astro-ph.CO]].

\bibitem{nia}
N.~Afshordi and J.~Magueijo,
Phys. Rev. D \textbf{94}, no.10, 101301 (2016)
doi:10.1103/PhysRevD.94.101301
[arXiv:1603.03312 [gr-qc]].

\bibitem{nia1}
M.~Mylova, M.~Moschou, N.~Afshordi and J.~Magueijo,
JCAP \textbf{07}, no.07, 005 (2022)
doi:10.1088/1475-7516/2022/07/005
[arXiv:2112.08179 [hep-th]].

\bibitem{kodama}
H. Kodama.
\newblock {Holomorphic Wave Function of the Universe}.
\newblock \emph{Phys.\ Rev.\ D}, 42:\penalty0 2548--2565, 1990.
\newblock \doi{10.1103/PhysRevD.42.2548}.

\bibitem{realkod}
J.~Magueijo,
Phys. Rev. D \textbf{104}, no.2, 026002 (2021)
doi:10.1103/PhysRevD.104.026002
[arXiv:2012.05847 [gr-qc]].


\bibitem{stephkod}
S.~Alexander, L.~Freidel and G.~Herczeg,
[arXiv:2212.07446 [hep-th]].

\end{thebibliography}
\end{document}